\documentclass[prd,aps,reprint,superscriptaddress,nofootinbib]{revtex4-2}

\pdfoutput=1

\usepackage[colorlinks=true
,urlcolor=blue
,anchorcolor=blue
,citecolor=blue
,filecolor=blue
,linkcolor=blue
,menucolor=blue
,linktocpage=true
,pdfproducer=medialab
,pdfa=true
]{hyperref}
\usepackage[T1]{fontenc}
\usepackage{fontawesome}
\usepackage{feynmf}
\usepackage{graphicx}
\usepackage{enumitem}
\usepackage{latexsym}
\usepackage{amsfonts}
\usepackage{amssymb}
\usepackage{mathrsfs}
\usepackage{color}
\usepackage{amsmath}
\usepackage[capitalise]{cleveref}
\usepackage{slashed}
\usepackage{dcolumn}
\usepackage{verbatim}
\usepackage{comment}
\usepackage{float}
\usepackage{multirow}
\usepackage{xspace}
\usepackage[normalem]{ulem}
\usepackage[dvipsnames]{xcolor}

\definecolor{triumfgreen}{HTML}{1D8038}
\definecolor{triumfred}{HTML}{F2705D}

\def\triumf{TRIUMF, 4004 Wesbrook Mall, Vancouver, BC V6T 2A3, Canada}
\def\sfu{Department of Physics, Simon Fraser University, Burnaby, BC V5A 1S6, Canada}

\begin{document}

\title{Hadronic Probes of Non-Standard Neutrino Interactions}


\author{Carlos Henrique de Lima}
\email{cdelima@triumf.ca}
\affiliation{\triumf}

\author{David McKeen}
\email{mckeen@triumf.ca}
\affiliation{\triumf}

\author{John Ng}
\email{misery@triumf.ca}
\affiliation{\triumf}

\author{Douglas Tuckler}
\email{dtuckler@triumf.ca}
\affiliation{\triumf}
\affiliation{\sfu}

\begin{abstract}
In this work, we study leptonic decays of hadrons as probes of light neutrinophilic scalars that mediate enhanced neutrino self-interactions. Such scalars can be emitted in processes involving neutrinos, turning two-body decays into three-body final states and producing characteristic spectral distortions. We compute these effects for charged pion decay and nuclear electron capture decay, including both on-shell and off-shell scalar emission, as well as the loop-induced renormalization required to cancel divergences. Using these results, we derive the projected sensitivity of PIONEER and assess the current and future reach of BeEST. The resulting low-energy spectral tails provide a characteristic signal for light neutrinophilic scalars, making upcoming hadron decay experiments powerful probes of light mediators of non-standard neutrino self-interactions.
\end{abstract}

\maketitle

\section{Introduction}\label{sec:intro}

Neutrinos are the most elusive particles in the Standard Model (SM), making it incredibly challenging to measure their properties experimentally. The SM predicts neutrino self-interactions that are far beyond the sensitivity of current and future experiments, leaving open the possibility that neutrino self-interactions are much stronger than the weak interaction. Indirect measurement of $Z$-neutrino coupling at LEP does not rule out strong neutrino self-interactions via the exchange of a new force carrier. 

New forces that act predominantly on SM neutrinos appear in many well-motivated extensions of the SM. They are present in, e.g., neutrino mass models~\cite{Chikashige:1980ui,Gelmini:1980re,Schechter:1981cv,Barger:1981vd,deLima:2022dht} and sterile neutrino dark matter models~\cite{DeGouvea:2019wpf,Kelly:2020pcy,Kelly:2020aks,An:2023mkf}. Furthermore, neutrino self interactions generate unique signatures at high-energy colliders~\cite{deLima:2024ohf,deGouvea:2019qaz,Dev:2021axj,Agashe:2024owh}, low energy experiments~\cite{Pasquini:2015fjv,Berryman:2018ogk,Kelly:2019wow,Brdar:2020nbj,Deppisch:2020sqh}, astrophysical processes~\cite{Kolb:1987qy,Ng:2014pca,Shoemaker:2015qul,Heurtier:2016otg,Das:2017iuj,Kelly:2018tyg,Bustamante:2020mep,Esteban:2021tub,Chang:2022aas,Chen:2022kal,Fiorillo:2022cdq,Fiorillo:2023ytr,Fiorillo:2023cas,Telalovic:2024cot}, and cosmology~\cite{Cyr-Racine:2013jua,Archidiacono:2013dua,Huang:2017egl,Escudero:2019gvw,Barenboim:2019tux,Blinov:2019gcj,Lyu:2020lps,Camarena:2024daj}.

A large class of models can be parametrized by the introduction of a light neutrinophilic singlet scalar $\phi$~\cite{DeGouvea:2019wpf,Kelly:2019wow,Kelly:2020aks,Kelly:2021mcd,Benso:2021hhh,Chichiri:2021wvw,Heeck:2017wgr,McKeen:2018xyz,Heeck:2019guh}. These interactions can derive from an dimension-5 Weinberg-like operator~\cite{Weinberg:1979sa} induced by the integration of additional heavy states above the electroweak symmetry scale.

A distinguishing feature of this class of models is the modification of weak-interaction processes to include the emission of a neutrinophilic scalar, increasing the final-state particle multiplicity in scatterings and decays involving neutrinos. This feature can be exploited in processes where the SM produces a two-body final state, since scalar emission increases this to a three-body final state and significantly modifies the energy distribution of the visible final states.

In this paper, we study two decays involving neutrinos that are primarily two-body: the leptonic decays of mesons and the electron capture decays of nuclei. In both cases, the emission of a neutrinophilic scalar leads to three-body kinematics, significantly modifying the energy distribution of visible particles in the final state (and amusingly echoing one of the original reasons for introducing the neutrino in the first place). For concreteness, we focus on two proposed and ongoing experiments targeting these processes, namely PIONEER~\cite{PIONEER:2025idw} and BeEST~\cite{Leach:2021bvh}. 

PIONEER will measure pion decays with an order-of-magnitude improvement in precision over current experiments~\cite{PiENu:2015seu}. The experimental setup allows for precise measurement of differential distributions of the charged lepton energy spectrum. At lower energies, the BeEST experiment measures $^7$Be electron-capture decays, and the precise measurement of the kinetic energy of the outgoing nucleus can shed light on the emission of a neutrinophilic scalar.

These decay processes are a standard means to search for massive (mostly) sterile neutrinos, where the two-body nature creates a clear peak in the energy distributions at a lower-than-expected energy. In the case of neutrinophilic scalar emission, the process becomes three-body, and thus, instead of a peak, there is a lower energy tail. A rough correspondence between the sensitivities in the two classes of searches is $U^2\sim\lambda^2/(32\pi^2)$, where $U$ is the active-sterile mixing angle of a heavy neutrino. The factor of $32\pi^2$ in the scalar case is due to the increased phase space. 

The three-body phase space nevertheless generates an advantage for the search for this class of beyond-the-SM (BSM) extensions. While sterile neutrino searches lose sensitivity for light masses, where the SM and BSM processes become indistinguishable, the same is not true for the three-body process, where at light masses a low-energy tail is present, significantly differing from the SM expectation. In addition, the emission of very light scalars in both processes we consider suffers from an apparent divergence. We demonstrate that these divergences are canceled in UV-complete models when including one-loop corrections to the two-body processes, leaving behind a mild model dependence.

The organization of this paper is as follows. In Section~\ref{sec:model}, we introduce the neutrinophilic scalar model and discuss UV completions. In Section~\ref{sec:nuscalar} we determine the effect of those scalars in pion decay experiments and estimate the reach of PIONEER for electron-neutrino coupled scalars. In Section~\ref{subsec:beest} we study the modifications to electron capture decay spectra induced by these scalars and present the current and future reach of the BeEST experiment. We conclude in Section~\ref{sec:conc}.

\section{Neutrinophilic Scalar Model and Existing Constraints}\label{sec:model}

Neutrinos may experience interactions far stronger than those predicted by the SM while remaining consistent with current terrestrial and astrophysical constraints. A simple possibility involves a light, gauge-singlet scalar $\phi$ that couples directly to neutrinos,
\begin{equation}
\label{eq:NSI}
\mathcal{L} \supset \frac{\lambda_{\alpha\beta}}{2} \phi \, {\bar{\nu}^c}_\alpha \nu_\beta + \text{h.c.} \, ,
\end{equation}
where $\alpha$ and $\beta$ label lepton flavors, allowing for stronger-than-weak neutrino self-interactions.\footnote{While we treat the neutrinos as essentially massless in this work and consider only the coupling of the scalar only to (active) left-handed neutrinos, this interaction could involve sterile neutrino degrees of freedom, which would be necessary if the light neutrinos were to be Dirac particles. This would not change our conclusions appreciably, so we focus on this more minimal case where the sterile degrees of freedom are parametrically heavier, and the light neutrinos are Majorana particles.} Such scalars can arise in models where lepton number is spontaneously broken, as in Majoron scenarios. The coupling in general can be non-diagonal in lepton flavor. For this work, when we analyze scenarios with $\lambda_{\ell \ell'}$, we assume the dominance of those couplings over others in the $\lambda$ matrix. The specific bound for the most general case with all couplings non-zero can be obtained with a proper rescaling of the bounds derived in this work. We assume that $\phi$ is complex, although the bounds we derive do not critically depend on this assumption and hold if $\phi$ is real.

Above the electroweak scale, the effective coupling can originate from the gauge-invariant, higher-dimensional operator
$g_{\alpha\beta} \phi (\bar L_\alpha^c H^\dagger)(L_\beta H)/\Lambda^2$, where $\Lambda$ is the scale of new physics, $L_\alpha = (\nu_\alpha, \ell_{L\alpha})^T$ is the SM lepton doublet, $H$ is the Higgs doublet, and $\lambda_{\alpha\beta} = g_{\alpha\beta} v^2/\Lambda^2$ with $v = 246~\mathrm{GeV}$. The Weinberg-like operator serves as the starting point for many phenomenological studies of light neutrinophilic scalars. An important point to note is that the low energy couplings in Eq.~\eqref{eq:NSI} are not gauge invariant, as emphasized in~\cite{Zhang:2024meg,Foroughi-Abari:2025mhj}. The heavy particles integrated out to generate these low energy interactions play a crucial role in rendering electroweak processes free from UV and IR divergences at the one-loop level.

A UV complete model that generates such an interaction at low energy is given by the inclusion of a heavy complex $SU(2)_L$ triplet with unit hypercharge, which is integrated out at a scale $\Lambda$~\cite{Berryman:2018ogk,Zhang:2024meg,Foroughi-Abari:2025mhj}. For this work, we have such UV completion in mind, but keep most results as model independent as possible. When produced experimentally, $\phi$ decays to neutrinos through the interaction in Eq.~(\ref{eq:NSI}) so that it appears as missing energy.

\section{Neutrinophilic scalars in pion decay}\label{sec:nuscalar}

\subsection{SM Pion Decays}\label{subsec:SM}

To identify potential new physics in pion decays, a precise understanding of the Standard Model (SM) process is essential, both theoretically and experimentally. While remaining general in our treatment, for definiteness we focus on the PIONEER experiment~\cite{PIONEER:2025idw}, the successor to PIENU~\cite{PiENu:2015seu}, which will offer unprecedented precision, at the ${\cal O}(0.01\%)$ level, in measuring the ratio of pion decay rates to an electron and muon, $R_{e/\mu}\equiv\Gamma_{\pi^+\to e^+\nu_e (\gamma)}/\Gamma_{\pi^+\to \mu^+\nu_\mu(\gamma)}$. To achieve this goal, the positron spectrum in this decay must be characterized with great precision. 

In the SM, the dominant process is the two-body decay $\pi^+ \to \ell ^+ \nu_\ell$ with a total decay width, at leading order, given by
\begin{equation}
\Gamma_{\pi^+\to \ell^+\nu_{\ell}} = \frac{G_\beta^2}{8 \pi} f_\pi^2 m_{\pi}^3 \delta_\ell \left(1 - \delta_\ell\right)^2\equiv \Gamma^0_{\ell} \, ,
\end{equation}
where $m_\pi$ is the charged pion mass, $\delta_\ell=m_{\ell}^2/m_{\pi}^2$, $f_\pi\simeq 130~\rm MeV$ is the pion decay constant~\cite{FlavourLatticeAveragingGroupFLAG:2024oxs}, and $G_\beta=G_FV_{ud}$ is the effective Fermi constant relevant for beta decay. In this process, in the $\pi^+$ rest frame, the charged lepton is monochromatic with energy $E_\ell=(m_\pi/2)(1+\delta_\ell)$. The helicity suppression in this decay mode famously renders the muon final state by far the dominant one, with the electronic channel down by $\sim10^4$. The ratio of the decay rates is calculated in the SM to be~\cite{Marciano:1993sh,Finkemeier:1995gi,Cirigliano:2007xi}.
\begin{equation}
R_{e/\mu}^{\rm{SM}}\equiv\frac{\Gamma_{\pi^+\to e^+\nu_{e}(\gamma)}}{\Gamma_{\pi^+\to \mu^+\nu_{\mu}(\gamma)}}=(1.2352 \pm 0.0001)\times 10^{-4} \, ,
\end{equation}
where the $(\gamma)$ in the processes indicates that QED corrections involving real and virtual photons have been taken into account.

The signal of this rare decay mode comes from observing a pion that is stopped inside a detector, then decays to a lepton carrying monochromatic energy. However, to achieve ${\cal O}(0.01\%)$ precision in the measurement of the positron channel requires measuring the spectrum at lower energies, which is populated by SM radiative corrections and detector effects. As we will see below, the emission of a neutrinophilic scalar leads to positrons in this lower energy range, and therefore, we need to understand the contributions to the low energy part of the spectrum to estimate the experimental reach.

The dominant radiative corrections come from the emission of virtual and real photons below the detector threshold. These corrections add a low energy tail to the positron energy distribution below the tree-level value. To leading order, these contributions come from so-called inner bremsstrahlung processes (emission from the initial and final charged states), structure-dependent contributions (emission from the hadron current), and the interference between them. These processes have been well understood for over half a century~\cite{Neville:1961zz,Brown:1964zza,DeBaenst:1968wha,Bryman:1982et,Holstein:1986uj,Bijnens:1996wm,Bryman:2025pet}.

An additional contribution to the positron energy distribution at energies below $m_\pi/2$ comes from the detector response to electromagnetic energy injection, which, even in the most idealized detectors, is not always fully contained. This response is challenging to estimate without a full detector simulation. As we focus on the PIONEER experiment, we utilize preliminary signal distributions found in Ref.~\cite{PIONEER:2025idw} to account for detector effects. 

To accurately estimate the reach for BSM physics, it is also necessary to understand the expected backgrounds that arise from other processes. In the case of $\pi^+\to e^+$ decays, these are mainly muons, either from the beam or produced by pions decaying in the target (since this decay channel is much more common). This is where PIONEER significantly improves upon its predecessor, PIENU. The experiment stops low energy pions in an active target, which is used for tracking with good vertex reconstruction. This makes it possible to distinguish prompt $\pi^+ \rightarrow e^+ \nu_e$ decays from the $\pi^+ \rightarrow \mu^+ \rightarrow e^+$ chain, where the muon survives for a few microseconds before decaying. Together with a high resolution calorimeter, PIONEER suppresses the background by several orders of magnitude while keeping the SM signal efficiency high.  There will still be a surviving background at PIONEER coming from muon decays in flight. The extent to which this can be suppressed is not yet known, and we model this uncertainty in our analysis at the end of this section. 

We also note that there are processes in the SM that are formally irreducible backgrounds to the signal that we consider, the second-order weak decays $\pi^+ \rightarrow \ell^+\nu_l \nu_{\ell^\prime} \bar{\nu}_{\ell^\prime}$. Such a process can mimic the signal of neutrinophilic scalars as they also generate non-trivial missing energy distributions for the visible lepton energy. However, the rates for these processes in the SM are incredibly small, with ${\rm Br}(\pi^+\rightarrow\mu^+ \nu_{\mu} \nu \bar{\nu})= 4 \times 10^{-20}$ and ${\rm Br}(\pi^+\rightarrow e^+ \nu_{e} \nu \bar{\nu})= 1.7 \times 10^{-18}$~\cite{Bardin:1970wq,Gorbunov:2016tbk}. 

These processes were out of reach of PIENU, which placed an upper bound of  ${\rm Br}(\pi^+\rightarrow\mu^+ \nu_{\mu} \nu \bar{\nu})<8.6\times 10^{-6}$ and ${\rm Br}(\pi^+\rightarrow e^+ \nu_{e} \nu \bar{\nu})<1.7\times 10^{-7}$~\cite{PIENU:2020las}. While PIONEER will significantly improve on these bounds, they will remain far beyond experimental reach and do not contribute to the background for our analysis. In the far future, measuring this SM process would mark an important milestone for neutrino physics, and contribute to a ``neutrino floor'' in the search for BSM neutrino self-interactions that are comparable to or weaker than the weak interaction. With the SM process under control, let us introduce the theoretical prediction for the neutrinophilic scalar.

\subsection{Neutrinophilic Scalars in Pion Decays}\label{subsec:nuscalar}

In this section, we discuss the modification of the rate for the pion to decay to a charged lepton and missing energy in the presence of a neutrinophilic scalar. We are as model-agnostic as possible here, focusing on the low energy observables. However, since the neutrinophilic interaction comes from a nonrenormalizable operator, there is a slight UV dependence in the decay rates. Additionally, we confront the divergences that arise when the scalar is parametrically light and show that they are properly canceled when including one-loop corrections. 

The decay of a pion to a charged lepton and missing energy was searched for by the PIENU collaboration, both in the case of three-body final states~\cite{PIENU:2021clt} and $\ell^+3\nu$~\cite{PIENU:2020las} final states. The former search maps to the on-shell production of a neutrinophilic scalar~\footnote{The model explored in~\cite{PIENU:2021clt} has a similar lepton energy distribution but does not suffer from the same divergences because it comes from a sterile neutrino scenario explored in~\cite{Batell:2017cmf}.} while the latter is directly applicable to the case where the scalar is too heavy to be produced on-shell. We discuss the predictions for each of these cases in turn below.

When $m_\phi< (m_\pi-m_\ell)$ it can be produced on-shell in the decay $\pi^+\to \ell^+\nu_{\ell'} \phi$ with a differential decay rate summing over neutrino flavors of
\begin{equation}
\label{eq:dgammadx}
\begin{aligned}
\frac{1}{\Gamma^0_{\ell}}&\frac{d\Gamma_{\pi^+\to \ell^+\nu\phi}}{dx_\ell}=\sum_{\ell^\prime}\frac{\left|\lambda_{\ell \ell '}\right|^2}{32\pi^2}\frac{\sqrt{x_\ell^2-4\delta_\ell}}{\delta_\ell\left(1-\delta_\ell\right)^{2}}
\\
&\times\left[x_\ell(1-x_\ell)+2\delta_\ell\right]\frac{\left(1+\delta_\ell-\delta_\phi-x_\ell\right)^2}{\left(1+\delta_\ell-x_\ell\right)^3}\, .
\end{aligned}
\end{equation}
where the lepton energy is $E_\ell=m_\pi x_\ell/2$ which runs over $2\sqrt{\delta_\ell}<x_\ell<1+\delta_\ell-\delta_\phi$ with $\delta_\phi\equiv m_\phi^2/m_\pi^2$. As $m_\phi \rightarrow 0$, the rate diverges logarithmically in the high-energy regime,
\begin{equation}
\begin{aligned}
\Gamma_{\pi^+\to \ell^+\nu\phi}&=\Gamma^0_{\ell}\times\sum_{\ell^\prime}\frac{\left|\lambda_{\ell \ell '}\right|^2}{32\pi^2}\log\frac{m_{\pi}^2}{m_\phi^2} \\
&\quad+({\rm finite~as~}m_\phi\to0) \, .
\end{aligned}
\end{equation}
This divergence is an artifact of the truncation in perturbation theory, and it is tamed by the inclusion of higher order corrections. At energies well below the cutoff of the UV theory, $\Lambda$, the interactions are neutrinophilic, and thus at one-loop the charged current is shifted only by the renormalization of the neutrino wave function~\cite{Zhang:2024meg,Foroughi-Abari:2025mhj}. This, in turn, generates a finite correction to the charged current coupling to lepton flavor $\ell$ that depends on the cutoff of the UV completion and the neutrinophilic scalar mass as
\begin{equation}
\begin{aligned}
    g_W^{R} &= g_W\left[1-\sum_{\ell^\prime}\frac{\left|\lambda_{\ell \ell '}\right|^2}{64\pi^2}\left( \log\frac{\Lambda^2}{m_\phi^2}+2\xi_{UV} \right) \right]
    \\
    &\equiv g_W+\delta g_W \, ,
\end{aligned}
\label{eq:delta_gW}
\end{equation}
where $\xi_{UV}$ is a model-dependent constant normalized so that in the triplet model $\xi_{UV}=1$. The logarithm is universal and related to the apparent UV divergence in the low-energy interaction. The shift of the two-body decay rate is then
\begin{equation}
\begin{aligned}
\delta\Gamma(\pi^+ \to \ell^+ \nu_\ell) &= \frac{2\delta g_W}{g_W}\Gamma^0_{\ell} 
\\
&= -\sum_{\ell^\prime}\frac{\left|\lambda_{\ell \ell '}\right|^2}{32\pi^2} \left(\log\frac{\Lambda^2}{m_\phi^2} + 2\xi_{UV} \right)\Gamma^0_{\ell} \, .
\end{aligned}
\end{equation}
Crucially, the one-loop correction of the two-body process cancels the $m_\phi\to 0$ divergence. What remains is a logarithmic dependence on the UV scale and a mild model dependence parameterized by $\xi_{UV}$. The full rate for a pion to decay to a charged lepton and missing energy can be written at one-loop in the neutrinophilic coupling as 
\begin{equation}\label{eq:supp}
\begin{aligned}
&\frac{d\Gamma_{\pi^+\to \ell^++{\rm inv.}}}{dE_\ell}=\frac{2}{m_\pi} \frac{d\Gamma_{\pi^+\to \ell^+\nu_{\ell'}\phi}}{dx_\ell}\Bigg|_{x_\ell=2 E_\ell/m_\pi}
\\
&\quad\quad+\Gamma^0_{\ell}\left[1-\sum_{\ell^\prime}\frac{\left|\lambda_{\ell \ell '}\right|^2}{32\pi^2} \left(\log\frac{\Lambda^2}{m_\phi^2} + 2\xi_{UV} \right)\right]
\\
&\quad\quad\quad\quad\times\delta\left(E_\ell-\frac{m_\pi^2+m_\ell^2}{2m_\pi}\right) \, ,
\end{aligned}
\end{equation}
which is manifestly finite as $m_\phi\to0$. As discussed in Sec.~\ref{subsec:SM}, the energy distribution of the two-body component of this decay is modified from the purely monochromatic form in this expression by QED corrections as well as by detector effects. Note that we do not include QED corrections to the neutrinophilic boson emission case itself since they amount to a relative correction of roughly ${\cal O}(10\%)$.

\begin{figure}[t!]
    \centering
    \includegraphics[width=0.99\linewidth]{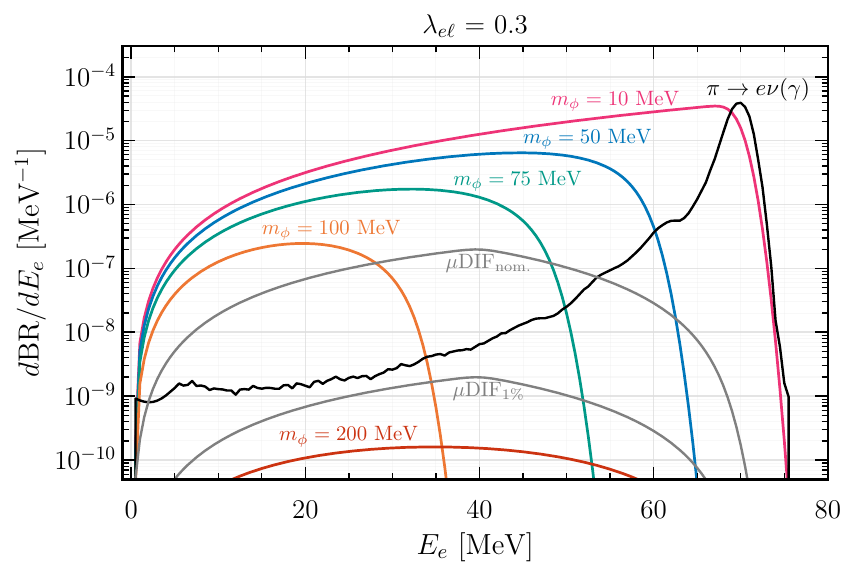}
    \caption{Differential branching fraction of the positron energy distribution for the SM (black), muon decay-in-flight background (gray), and the neutrinophilic scalar with $\lambda_{e
\ell} = 0.3$ for different $\phi$ masses. We use two $\mu$ decay-in-flight background hypotheses (gray): (i) $\mu\text{DIF}_{\text{nom.}}$ which is what is obtained with an average $\mu$ stopping time of $12~{\rm ps}$ and no further suppression and (ii) $\mu\text{DIF}_{\text{1\%}}$ where this background is suppressed to 1\% of $\mu\text{DIF}_{\text{nom.}}$, which aligns with the precision target of PIONEER. See text for details.}
    \label{fig:piondist}
\end{figure}

We show the positron energy spectrum in pion decay to neutrinophilic scalars of mass $m_\phi=10$, $50$, $75$, and 100 MeV coupled to the electron neutrino with $\lambda_{ee}=0.3$ in Fig.~\ref{fig:piondist}. The branching ratio to a neutrinophilic scalar does not suffer from chirality suppression as in the two-body decay, even though it involves a higher-body phase space. As a result, it can be comparable to the $e^+\nu_e$ final state for $\lambda_{ee}$ of ${\cal O}(m_e/m_\pi)$.

For $m_\phi> (m_\pi-m_\ell)$, the neutrinophilic scalar cannot be produced directly in pion decays, and the leading correction to the decay to a charged lepton and missing energy is the four-body process  $\pi^+\to \ell^+\nu_\ell \phi^* \rightarrow  \ell^+\nu_\ell \nu_{\ell'}\nu_{\ell'}$. Assuming that the $\phi$ coupling is diagonal in lepton flavor, the differential decay width to neutrinos of the same family has a simple form in the regime where $m_{\pi}\ll m_\phi<\Lambda$,
\begin{equation}
\begin{aligned}
&\frac{1}{\Gamma^0_{\ell}}\frac{d\Gamma_{\pi^+\to\ell^+ 3\nu_{\ell}}}{dx_\ell}=\frac{\left|\lambda_{\ell \ell}\right|^4}{4(4\pi)^4} \frac{m_\pi^4}{ m_\phi^4} \frac{\sqrt{x_\ell^2-4\delta_\ell}}{\delta_\ell\left(1-\delta_\ell\right)^{2}}\\
&\times\Big[x_\ell(1-x_\ell)+2\delta_\ell\Big]\Big(1+\delta_\ell-x_\ell\Big) \, ,
\end{aligned}
\end{equation}
where the lepton energy varies from $2\sqrt{\delta_\ell}<x_\ell<1+\delta_\ell$. The partial width into other lepton flavors is related to this by a factor of $\left|\lambda_{\ell^{'} \ell^{'}}\right|^2/(12  \left|\lambda_{\ell \ell}\right|^2)$ with the $1/12$ due to interference. This rate has the same distribution as in the case of a heavy neutrinophilic vector in Ref.~\cite{Bardin:1970wq} with an appropriate rescaling by the effective coupling strength of the four-Fermi interaction. The differential distribution in the case of pion decay to a positron can be seen in Fig~\ref{fig:piondist} for $m_\phi=200$ MeV, using the same detector response as in the on-shell $\phi$ case.

In this large $m_\phi$ regime, constraints can also be placed based on the modification of the two-body pion decay rates induced at loop level. In the presence of an electron-neutrino--specific coupling, the rates for $\pi^+\to e^+\nu_e$ and $\pi^+\to\mu^+\nu_\mu$ are modified by different shifts at one-loop to the weak charged current coupling in Eq.~(\ref{eq:delta_gW}),
\begin{equation}\label{eq:ratio}
\begin{aligned}
R_{e/\mu}= R_{e/\mu}^{\rm SM}\left[1-\frac{\left|\lambda_{ee}\right|^2}{32\pi^2}\left( \log\frac{\Lambda^2}{m_\phi^2}+2\xi_{UV} \right) \right]\, .
\end{aligned}
\end{equation}
The resulting bound on $\lambda_{ee}$ has mild model dependence. In what follows we choose the triplet model of Refs.~\cite{Berryman:2018ogk,Zhang:2024meg,Foroughi-Abari:2025mhj} so that $\xi_{UV}=1$ with a cutoff of $\Lambda=500~\rm GeV$.

Armed with both the SM and BSM predictions computed above, we estimate the PIONEER experiment reach for neutrinophilic scalars produced in pion decay below.\footnote{As an aside, we note that the existence of a neutrinophilic scalar would also modify the Michel spectrum in muon decay. However, since muon decay is already three-body in the SM, the effect of adding another particle in the final state is much more modest than in the case of pion decays, where there is a qualitative difference in going from a two- to three-body final state, leading to weaker constraints. In addition, the overall $\mu$ decay rate could be shifted due to $\phi$ emission from its SM value as predicted by electroweak observables and potentially also lead to apparent violations of CKM unitarity. Ref.~\cite{Bryman:2019bjg} studied this in detail, showing that exotic muon branchings could be limited to the $10^{-3}$ level corresponding to coupling values of $\lambda\gtrsim0.5$. For these reasons, we do not show the reach for neutrinophilic scalars from muon decays. Similar statements can be made for $\tau$ decays, where the shift of leptonic decay rates from their SM expectations can be limited to the $10^{-2}$ level~\cite{Bertoni:2014mva,Bryman:2021ilc}.}

\subsection{PIONEER Analysis}
To determine the reach of PIONEER to neutrinophilic scalars, we perform a log-likelihood ratio test assuming that each bin (which we take to be $1~{\rm MeV}$ in size over the entire range) of the positron energy distribution is an independent counting experiment obeying Poisson statistics. We take the SM distribution from Ref.~\cite{PIONEER:2025idw} and assume that $10^8$ $\pi\to e$ events are recorded. We further assume that the dominant background comes from $\mu$ decays-in-flight and compare the signal plus SM and background hypothesis to one consisting of just the SM and background. To approximate the detector response to the signal, we smear the positron energy distribution with a Gaussian energy resolution with a width given by the fractional positron energy uncertainty, which we vary linearly from 10\% at $E_e=10~{\rm MeV}$ to 1.5\% at $E_e = 70~{\rm MeV}$~\cite{PIONEER:2025idw}. 

For the $\mu$DIF background level we consider two benchmark possibilities: (i) the number of $\mu$DIF events is what is expected for muons produced in $\pi^+\to\mu^+\nu_\mu$ that decay before the stopping time in the target of $12~{\rm ps}$ (which is the approximate transit time assuming a stopping power of $dE/dx\sim 40~{\rm MeV/cm}$ in the silicon active target for a $4.1~{\rm MeV}$ muon) and (ii) that the $\mu$DIF background is suppressed to 1\% of this nominal value. We show these two background levels in Fig.~\ref{fig:piondist} as gray curves labeled ``$\mu{\rm DIF}_{\rm nom.}$'' and ``$\mu{\rm DIF}_{1\%}$'' respectively. $\mu{\rm DIF}_{1\%}$ represents approximately the targeted level of background suppression at PIONEER, while $\mu{\rm DIF}_{\rm nom.}$ allows us to study the reach for a neutrinophilic scalar if this target cannot be reached.

\begin{figure*}[tbh!]
    \centering
    \includegraphics[width=0.77\linewidth]{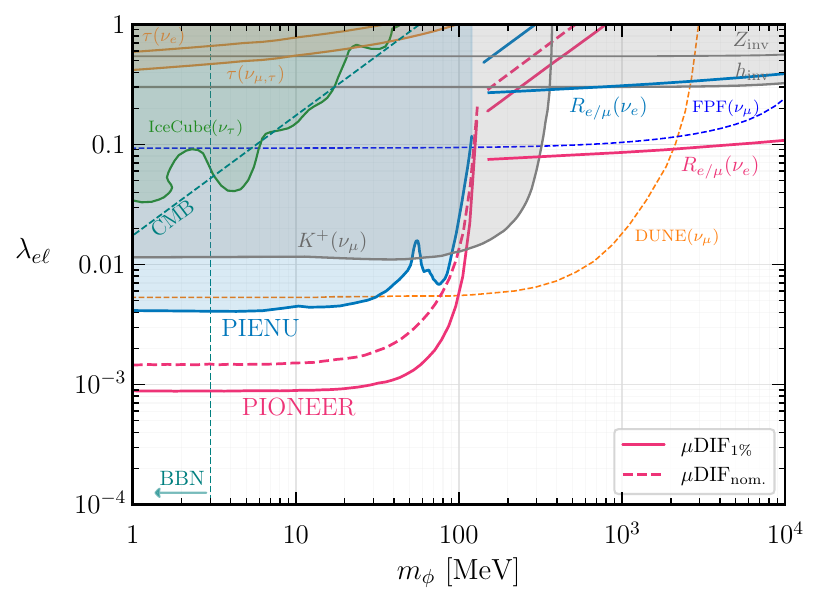}
    \caption{Projected $95\%$~CL sensitivity of PIONEER to neutrinophilic scalars coupled to electron neutrinos, shown in the $(m_\phi,\lambda_{e\ell})$ plane with $\Lambda=500$ GeV and $\xi_{UV}=1$. The solid (dashed) pink curve corresponds to the $\mu\text{DIF}_{\text{1\%}}$ ($\mu\text{DIF}_{\text{nom.}}$) background assumptions. Existing laboratory constraints from PIENU, kaon decays, invisible $Z$ and Higgs decays are shown for comparison, together with indicative constraints from neutrino experiments and cosmology. The shaded teal region indicates the parameter space favored by recent CMB analyses of enhanced neutrino self-interactions, while the vertical teal line shows the approximate lower bound from BBN, assuming standard cosmology. We also show the future sensitivity for DUNE~\cite{Kelly:2019wow} and the Forward Physics Facility (FPF) \cite{Kelly:2021mcd} for comparison.}
    \label{fig:pioneerbound}
\end{figure*}

The resulting 95\% CL expected upper limits on the coupling $\lambda_{e\ell}$ for $\ell=e$, $\mu$, $\tau$ (assuming only one coupling dominates in each case) are shown as a function $m_\phi$ in Fig.~\ref{fig:pioneerbound}. The solid pink curve depicts the reach with the $\mu$DIF background suppressed to the $\mu{\rm DIF}_{1\%}$ level, while the dashed pink curve shows the reach if it remains at the $\mu{\rm DIF}_{\rm nom.}$ level. In either case, PIONEER probes new regions of parameter space for neutrinophilic scalars with couplings $\lambda_{e\ell} \sim 10^{-3}$ for $m_\phi \lesssim 100~{\rm MeV}$. Additionally, PIONEER has sensitivity to off-shell neutrinophilic scalars with coupling $\lambda_{e\ell} \sim 0.3$ for $m_\phi \gtrsim 200~{\rm MeV}$. We also show current experimental constraints, which we explain further below.

\subsection{Existing Constraints}

Constraints on $\phi$ come from existing searches for invisible Higgs decays, invisible $Z$ boson decays, other charged meson decays, and cosmology. We briefly describe these here.

\subsubsection{Invisible Higgs Decays}
Invisible Higgs decays have long been a signal of BSM physics~\cite{Shrock:1982kd}. The Weinberg-like operator that generates the interaction in Eq.~\eqref{eq:NSI} also induces a neutrinophilic scalar coupling to the SM Higgs boson 
\begin{equation}
\mathcal{L} \supset \frac{\lambda_{\alpha\beta}}{v}  {\bar{\nu}^c}_\alpha \nu_\beta \,\phi \,h \,,
\end{equation}
which gives a new decay channel for the Higgs boson $h \to \nu_\alpha \nu_\beta \phi$ when $m_\phi < m_h$. Because $\phi$ decays invisibly, this decay contributes to the invisible width of the Higgs boson with a partial width given by
\begin{equation}
\Gamma(h \to \nu_\alpha \nu_\beta \phi) \simeq \frac{G_F  m_h^3 |\lambda_{\alpha\beta}|^2}{192 \sqrt{2}\pi^3}\,
\end{equation}
where we have taken the $m_\phi \ll m_h$ limit. Note, the rate has an additional factor of 1/2 for $\alpha = \beta$. The current upper limit on the Higgs invisible decay branching from ATLAS Run I and Run II combinations is BR$^h_\text{inv} < 0.107$~ \cite{ATLAS:2023tkt}, which translates to an upper limit on the neutrinophilic scalar coupling of
\begin{equation}
|\lambda_{\alpha\beta}| \lesssim 0.3.
\end{equation}
Future Higgs factories are expected to constrain the Higgs invisible branching ratio to less than 1\%, with the strongest expected reach from ILC and FCC-ee at 0.16\% and 0.19\%, respectively \cite{Blondel:2021ema,Potter:2022shg}. 

\subsubsection{Invisible $Z$ Decays}
A neutrinophilic scalar can have non-trivial effects in invisible $Z$ boson decays. For $m_\phi < m_Z$, $\phi$ can be radiated off the final state neutrinos, leading to the three-body decay $Z \to \nu_\alpha \nu_\beta \phi$. In addition, $\phi$ can modify electroweak couplings at loop order, as pointed out in \cite{Zhang:2024meg,Foroughi-Abari:2025mhj}. An upper limit on $\lambda_{e\ell}$ is obtained by requiring that the $\phi$ contribution to the $Z$ invisible decay width is within $2\sigma$ of the derived value $\Gamma(Z \to \text{inv.}) = 499.3 \pm 1.5$~MeV~\cite{ParticleDataGroup:2024cfk}. We show this for the flavor off-diagonal coupling in Fig.~\ref{fig:piondist}, which is constrained to $\lambda_{e\ell} \lesssim 0.5$. This bound is weaker by a factor of $\sqrt{2}$ for flavor diagonal couplings.

\subsubsection{Charged Meson Decays}
The predecessor of PIONEER, PIENU, has searched for a signal that is similar to the one explored in this work $\pi \rightarrow e \nu X$~\cite{PIENU:2021clt} (on-shell) and $\pi \rightarrow e 3\nu $~\cite{PIENU:2020las} (off-shell). The positron energy distributions are close enough to those of our signal that it is possible to reinterpret the PIENU bounds on branching ratios to the coupling $\lambda_{e\ell}$, which we show in  Fig.~\ref{fig:pioneerbound}. Note that the PIENU search for  $\pi \rightarrow e \nu X$ loses sensitivity around $m_{\phi}\simeq60$~MeV because the $\pi \rightarrow \mu \rightarrow e$ background has a similar distribution to a neutrinophilic scalar with that mass. This background is expected to be suppressed at PIONEER, and the surviving muon-decay-in-flight background is different enough that it is possible to differentiate from the signal, even if the search is background-dominated in that energy region.

The analysis performed focuses exclusively on the low energy tail, but it is also possible to look for deviations from the total decay widths. The effect depends on the specific flavor assignment. For example, a neutrinophilic scalar that couples only to electron neutrinos suppresses the theoretical prediction for the electronic decay as seen in Eq.~\eqref{eq:supp} on top of generating a low-energy tail. The current precision of $0.1\%$ and future PIONEER precision of $0.01\%$ on the $\pi^+ \rightarrow e^+ \nu_e (\gamma)$ allows for competitive bounds in the heavy mass regime of the mediator. 

An additional constraint on large $m_\phi$ is placed by considering modifications to $R_{e/\mu}$, as in Eq.~\eqref{eq:ratio}. PIENU measured this ratio to be $R_{e/\mu} = (1.2327\pm 0.0023) \times 10^{-4}$~\cite{PiENu:2015seu}, and we require that the loop induced $\phi$ contribution is within the $2\sigma$ uncertainty. This bound is shown in Fig.~\ref{fig:pioneerbound} by the blue curve labeled $R_{e/\mu} (\nu_e)$. PIONEER aims to decrease the uncertainty on $R_{e/\mu}$ by an order-of-magnitude. Assuming that PIONEER measures the SM value, we obtain the bound depicted by the pink curve labeled $R_{e/\mu} (\nu_e)$ by similarly requiring that the $\phi$ contribution is within the $2\sigma$ uncertainty. This effect is logarithmically sensitive to the cutoff scale $\Lambda$ and depends on the specific UV completion.

Sub-GeV neutrinophilic scalars can also be produced in leptonic decays of other charged mesons. The NA62 experiment performed the first search for invisible particles in the three-body decay $K^+ \to \mu^+ \nu_\mu X$ \cite{NA62:2021bji} where $X$ is an invisibly-decaying boson coupled to muons (and neutrinos, depending on the model), placing upper limits on the branching ratio for this process. Using an appropriately rescaled Eq.~\eqref{eq:dgammadx}, we reinterpret the NA62 bounds on the $K^+ \to \mu^+ \nu_\mu X$ branching ratio to place upper limits on the off-diagonal coupling $\lambda_{e\mu}$ of the neutrinophilic scalar as a function of $m_\phi$. This limit is shown in Fig.~\ref{fig:pioneerbound} by the gray curve labeled $K^+(\nu_\mu)$.

\subsubsection{Tau Lepton Decays}
Neutrinophilic scalars can be produced in the leptonic $\tau$ decays $\tau \to \ell \nu_\tau \nu_\ell \phi~(\ell = e, \mu)$, where $\phi$ is emitted from the final-state $\nu_\tau$ for nonzero $\lambda_{e\tau}$, or from the $\nu_\ell$ when $\lambda_{e\ell} \neq 0$ with $\ell \neq \tau$. Constraints on $\phi$ from these decays were derived in \cite{Brdar:2020nbj}, and are shown in Fig.~\ref{fig:pioneerbound} by the brown curves in the upper-left corner. These are typically much weaker than bounds from meson decays. 

\subsubsection{Astrophysical Neutrinos}
New neutrino self-interactions can affect the propagation of astrophysical neutrinos, modifying the observed neutrino spectrum in neutrino experiments like IceCube. Constraints on tau-neutrino coupled scalars are placed by studying the spectrum of IceCube high-energy starting events \cite{Esteban:2021tub}, and are shown by the green shaded region in Fig.~\ref{fig:pioneerbound}. Note, this constraint only applies for $\lambda_{e\tau}$, but is expected to be similar for other couplings.

\subsubsection{Cosmology}
New neutrino self-interactions can leave an imprint on cosmological observables, namely, big bang nucleosynthesis (BBN) and the cosmic microwave background (CMB). 

Neutrino self-interactions mediated by $\phi$ exchange delay the onset of neutrino free streaming, leading to modifications of the CMB power spectrum~\cite{Cyr-Racine:2013jua}. Recent analyses of the CMB power spectrum along with large scale structure data find an upper limit on the effective neutrino self-interaction strength at around the $\lambda^2/m_\phi^2=10^{-4}~{\rm MeV}^{-2}$ level~\cite{Camarena:2024daj,Poudou:2025qcx}. The 95\% upper limit from~\cite{Poudou:2025qcx} is depicted by the dashed teal line labeled ``CMB'' in Fig.~\ref{fig:pioneerbound}.

A light neutrinophilic scalar contributes to the effective number of relativistic degrees of freedom during BBN, which gives a lower bound $m_\phi \gtrsim$ MeV. This is shown in  Fig.~\ref{fig:pioneerbound} by the vertical dashed teal line.


\section{Neutrinophilic scalar Emission in  Electron Capture}\label{subsec:beest}

A related class of decays of hadronic systems involving the charged current weak interaction is the electron capture decay (EC) of unstable nuclei. In this process, an atomic electron is captured by a nucleus, inducing a transition to a nucleus of smaller charge along with a neutrino. This process is essentially two-body, meaning that the recoiling nucleus has a well-defined kinetic energy dictated by the $Q$-value of the reaction and the mass of the neutrino. The two-body nature of the final state can be used to measure the neutrino masses and to search for heavy, mostly sterile neutrinos~\cite{Filianin:2014gaa,Brodeur:2023eul} in a way that is very analogous to searches using meson decays. The $Q$-values available in EC decays are much smaller than the pion mass. This means, e.g., that such experiments are sensitive to sterile neutrino masses that are in the regime where pion decay experiments lose sensitivity since a sterile neutrino of such a mass is effectively degenerate with the active neutrinos and can therefore be decoupled from weak interactions.

In this section, we study the effects on EC induced by the existence of a light neutrinophilic scalar. As in the case of pion decays, the phase space of the final state goes from two-body to three-body, and the visible particle in the final state is no longer monochromatic. For definiteness, we focus on the ongoing BeEST experiment, which involves $^7{\rm Be}$ implanted into sensitive superconducting tunnel junction (STJ) radiation detectors~\cite{Friedrich:2022,Leach:2021bvh} that can measure the kinetic energy of the recoiling  $^7{\rm Li}$ atoms with ${\cal O}({\rm eV})$ precision. We start by discussing SM EC decays and then move on to estimate the reach that EC with $\phi$ emission could have. While the present reach for neutrinophilic scalars in EC decays is not terribly strong, scaling such experiments up could potentially probe interesting regions of parameter space in the future.

\subsection{Standard Electron Capture Decays}

To look for beyond-SM effects in EC decays, we first need to understand the process in the context of the SM. The reaction in question is
\begin{equation}
Ne^-\to N^\prime \nu_e\, ,
\label{eq:ECSM}
\end{equation}
where $N^{(\prime)}$ is the initial (final) state nucleus with mass $M^{(\prime)}$. The $Q$-value in this process is
\begin{equation}
Q=M+m_e-M^\prime\, ,
\end{equation}
where we have ignored the electron's atomic binding energy, which is a good approximation in the case of the light nuclei considered in this work. A Fermi or Gamow-Teller EC rate can be related to the nuclear matrix element through
\begin{equation}
\begin{aligned}
\Gamma_{\rm EC}&\simeq\frac{G_\beta^2}{2\pi } Q^2\left|\psi_e(0)\right|^2\frac{ \left| {\cal M} (Q)\right|^2}{(2J_i+1)A^2}\, ,
\end{aligned}
\end{equation}
where the electron wavefunction $\psi_e$ is evaluated at the position of the nucleus, $A$ is the number of nucleons, and
\begin{equation}
{\cal M}(Q)=\langle N^\prime|{\cal H}_{\rm nuc}|N\rangle 
\end{equation}
is the nuclear matrix element evaluated at momentum transfer $Q$ with ${\cal H}_{\rm nuc}$ the hadronic charged-current weak Hamiltonian.  In the case of a massless neutrino, the kinetic energy of the recoiling nucleus is
\begin{equation}
\begin{aligned}
T\simeq\frac{Q^2}{2M}\, .
\end{aligned}
\end{equation}
Importantly, the recoil of the nucleus is independent of the complicated nuclear matrix elements, even in the presence of neutrino masses, which shifts the kinetic energy downward by $m_\nu^2/2M$. The same feature occurs in the case of heavy sterile neutrinos that could be produced in EC, where the information of the nuclear matrix elements factors out, and the physical observable is the existence of a lower kinetic energy recoil peak.

\subsection{BeEST Experiment}

The BeEST experiment searches for exotic signatures of BSM physics in electron capture (EC) decay by measuring the nuclear recoil energy of the daughter nucleus with cryogenic superconducting tunnel junction (STJ) detectors operated at millikelvin temperatures. Radioactive $^{7}\mathrm{Be}$ with a half-life $\simeq 53$~days is implanted directly into the absorber, ensuring that the nuclear recoil energy from the decay is locally thermalized and measured with $\mathcal{O}(\mathrm{eV})$ energy resolution.

In the EC decay of $^{7}\mathrm{Be}$, the nucleus captures an orbital electron and decays to $^{7}\mathrm{Li}$, predominantly to the ground state ($\simeq 90\%$) with a $Q$-value of $\simeq 862~\mathrm{keV}$, and to an excited state ($\simeq 10\%$) with a $Q$-value of $\simeq 384~\mathrm{keV}$. These two-body decays produce monoenergetic nuclear recoil energies of approximately $56~\mathrm{eV}$ and $11~\mathrm{eV}$, respectively, in the absence of additional de-excitation processes. For decays to the excited state, the subsequent emission of a $477~\mathrm{keV}$ $\gamma$ ray imparts an additional recoil momentum to the $^{7}\mathrm{Li}$ nucleus. This leads to Doppler broadening of the recoil feature and contributes an additional $\sim 20~\mathrm{eV}$ of recoil kinetic energy, depending on the $\gamma$-ray emission direction. The $\gamma$-ray itself typically escapes the absorber, so that the dominant effect on the measured spectrum arises from the modified nuclear recoil rather than direct $\gamma$-ray energy deposition.

In practice, the measured recoil spectrum exhibits multiple effective peaks arising from the combination of nuclear recoil, atomic final-state effects, and solid-state energy deposition processes. In representative BeEST datasets, four prominent features appear near approximately $105$, $80$, $56$, and $30~\mathrm{eV}$. These are treated phenomenologically as effective peaks in the measured energy spectrum and are modeled using Gaussian components, as discussed in Ref.~\cite{BeEST:2024pfa}, rather than as fixed fundamental recoil energies. Finite energy resolution, atomic relaxation cascades, and incomplete containment of de-excitation products contribute to additional broadening and shape distortions of these features, which are empirically incorporated into the spectral modeling. The modeling of the BeEST experiment used in this work follows Ref.~\cite{BeEST:2024pfa}.

The nuclear matrix elements in the case of $^7{\rm Be}$ to both $^7{\rm Li}$ states are Gamow-Teller, so that in each case we can write
\begin{equation}
{\cal M}^i(Q)=\langle ^7{\rm Li}^{(\ast)}|g_A\sigma^i\tau_-|^7{\rm Be}\rangle \, ,
\end{equation}
where $g_A$ is the nucleon axial vector coupling, $\sigma^i$ is a spin operator, and $\tau_-$ is an isospin-lowering operator. The computation of the nuclear matrix element is highly non-trivial, but in reach of current nuclear theory calculations~\cite{Gysbers:2019uyb,Gennari:2024sbn}. The analysis we perform is always relative to the SM rate, and most of the nuclear and atomic dependence drops out. Nevertheless, having accurate nuclear predictions can be useful to open another avenue to probe BSM physics, where the overall normalization of the rates becomes a testable observable.

\begin{figure}[t]
    \centering
    \includegraphics[width=0.99\linewidth]{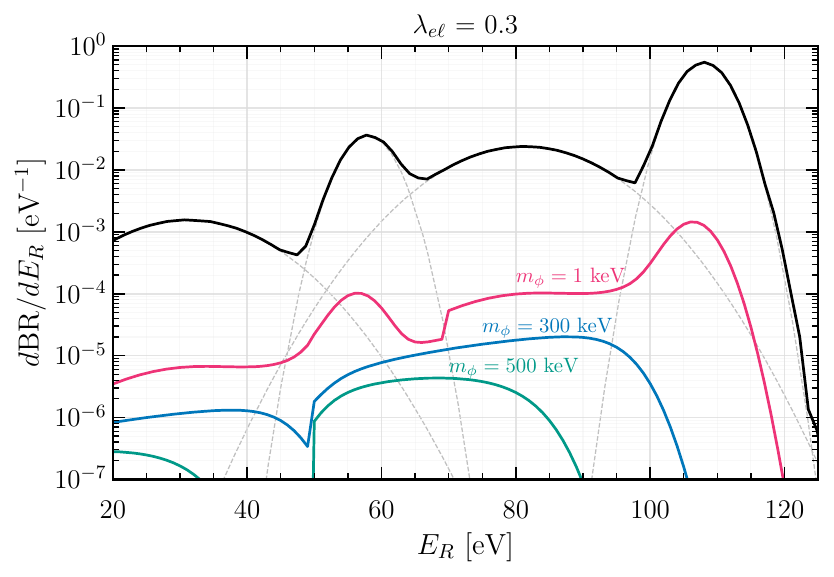}
    \caption{Differential Branching fraction of the recoil energy distribution for the SM (black) and the neutrinophilic scalar with $\lambda_{e\ell} = 0.3$. In dashed gray, we show the Gaussian fits of the four decay peaks used for the analysis.}
    \label{fig:BeESTspec}
\end{figure}

For our preliminary study, we simply model each of the four peaks with Gaussians with widths and heights determined by the experimental measurements, essentially normalizing the nuclear matrix elements and atomic wavefunctions to data. This is a simple modeling of the system, but captures the most essential features of the BSM contributions for this process, which comes from the existence of low energy tails at each monochromatic peak. In Fig.~\ref{fig:BeESTspec} we show our modeled spectra alongside the phenomenological fit done in Ref.~\cite{BeEST:2024pfa}. Our results agree reasonably well with those found in Ref.~\cite{BeEST:2024pfa}, giving us confidence that we can describe the BeEST data sufficiently well for our purposes. Given our description of the SM EC decays at BeEST, we move on to discuss the effect of the neutrinophilic scalar on the EC spectrum.

\subsection{Neutrinophilic scalar rate}\label{subsec:ECrates}

The process of interest now is the emission of $\phi$ by the outgoing neutrino in an EC decay,
\begin{equation}
Ne^-\to N^\prime \nu_\ell\phi\, ,
\label{eq:ECBSM}
\end{equation}
which can occur if $m_\phi<Q$. In this reaction, the kinetic energy of the final state nucleus lies between
\begin{equation}
0<T<\frac{Q^2-m_\phi^2}{2M}\, .
\end{equation}
Given the $\phi$ coupling to neutrinos in Eq.~\eqref{eq:NSI}, the differential rate for $\phi$ emission as a function of the kinetic energy of the recoiling nucleus can be related to that for standard EC, 
\begin{equation}
\begin{aligned}
\frac{1}{\Gamma_{\rm EC}}\frac{d\Gamma_{{\rm EC}\phi}}{dT}&=\frac{\left|\lambda_{e \ell}\right|^2}{8\sqrt2 \pi^2}\sqrt{\frac{M^3T}{Q^2}}
\\
&\quad\quad\times\frac{(Q^2-m_\phi^2-2MT)^2}{(Q^2-2MT)^3}\left|F(T)\right|^2\, .
\label{eq:GammaECphi}
\end{aligned}
\end{equation}
Forming this ratio eliminates the dependence on the atomic wavefunctions while putting the dependence of the nuclear matrix elements into a form factor,
\begin{equation}
\left|F(T)\right|^2\equiv\frac{|{\cal M}(\sqrt{Q^2-2MT})|^2}{\left|{\cal M}(Q)\right|^2}\, ,
\end{equation}
which is the ratio of the nuclear matrix elements evaluated at the momentum transfer relevant for either standard EC, $Q$, or EC with $\phi$ emission, $\sqrt{Q^2-2MT}$. In the case of $^7{\rm Be}$ decay, the $Q$-values are much smaller than any of the nuclear physics scales involved in the problem so that the matrix elements do not vary appreciably as the momentum transfer ranges below the value in standard EC, resulting in $F(T)=1$.\footnote{This fact is also crucial in extracting limits on heavy sterile neutrinos where it is assumed that the nuclear matrix elements at $Q$ and $(Q^2-m_N^2)^{1/2}$ are the same so that the rate of sterile neutrino production is simply modulated by the square of its mixing angle, $U^2$.} 

The emission of $\phi$ in the EC process produces a tail at lower recoil energies than in the standard, two-body case due to the increased phase space. Additionally, just as in the pion decay case, the total rate diverges as $m_\phi\to0$ near where the recoil energy is close to the standard EC value. In the small $m_\phi$ limit, the total rate is
\begin{equation}
\begin{aligned}
\frac{\Gamma_{{\rm EC}\phi}}{\Gamma_{\rm EC}}&\simeq\frac{\left|\lambda_{e\ell}\right|^2}{32\pi^2}\left(\log\frac{4Q^2}{m_\phi^2}-\frac72\right)
\end{aligned}
\end{equation}
Again, this divergence cancels with the divergent wavefunction renormalization of the two-body rate. 

One important thing to note here is that, unlike the $\pi\to e$ decay, the electron capture process is not chirality-suppressed. Therefore, the $\phi$ emission process for $\lambda\lesssim 1$ is at a much lower rate than the SM process being targeted, whereas in the $\pi$ decay case, it is comparatively less suppressed compared to the $\pi^+\to e^+\nu_e$ process.

As in the pion decay case, it is in principle possible to bound the $\phi$ coupling when $m_\phi$ is too large for it to be produced on-shell in EC by its one-loop effect on the rate for standard EC. In pion decays, this is possible for flavor non-universal couplings since forming the ratio of decays to $e\nu_e$ and $\mu\nu_\mu$ allows for QCD uncertainties to cancel. Since such a ratio cannot be formed in the case of EC decays, using the total rate to bound $\lambda$ would require understanding the nuclear and atomic physics to the ${\cal O}(10^{-3})$ level, which is a daunting task.

With all the ingredients we need ready, we perform a preliminary analysis for the search for neutrinophilic scalars at BeEST in the next section.

\subsection{BeEST analysis}\label{subsec:EClimits}
\begin{figure*}[tbh!]
    \centering
    \includegraphics[width=0.77\linewidth]{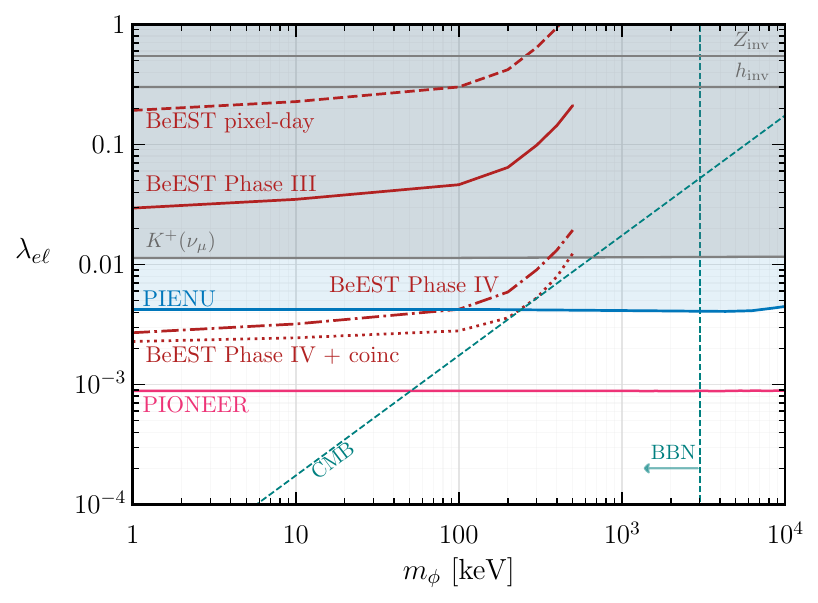}
    \caption{Projected sensitivity of BeEST to neutrinophilic scalars coupling to electron neutrinos. The curves correspond to increasing experimental exposure and capability: a single pixel-day, the current Phase~III configuration, a projected Phase~IV exposure, and Phase~IV with photon-coincidence tagging of decays to the excited nuclear state. Existing laboratory constraints and the projected PIONEER sensitivity are shown for comparison.}
    \label{fig:beestbound}
\end{figure*}
We perform an analysis looking for deformations of the SM two-body decay peak spectrum. For the SM signal, we model the recoil energy with four Gaussians for each of the possible transitions that best fit the experimental distribution measured in Ref.~\cite{BeEST:2024pfa}; the widths of each of the Gaussians are different, reflecting the different physics underlying the spread in each decay channel. For the $\phi$-emission signal, we then smear the recoil energies by the Gaussian corresponding to the underlying nuclear transition, generating the spectrum shown in Fig.~\ref{fig:BeESTspec} for $\lambda_{e e}=0.3$ and $m_\phi=1$, $300$, and $500~{\rm keV}$. We then bin both signal and background in $1~{\rm eV}$ bins starting at $20~{\rm eV}$ and form a Poisson log-likelihood for each $m_\phi$ hypothesis, similar to the analysis performed in the last section, and estimate expected upper bounds on the coupling. Sensitivity projections for future BeEST phases assume representative detector exposures (e.g., Phase~III corresponding to tens of pixels operated for $\mathcal{O}(50)$ days, and Phase~IV corresponding to $\mathcal{O}(100)$ pixels operated for $\mathcal{O}(100)$ days). Note that we do not include systematic or modeling floors on the event rate in determining our future sensitivities.

We consider four scenarios with corresponding estimated limits shown in Figure~\ref{fig:beestbound}. First, we estimate the reach using a single pixel over one day corresponding to around $10^6$ $^7{\rm Be}$ ``clean'' EC decays. Next, we scale this reach by 36 pixels over 50 days, which is the current phase (termed ``Phase III'') of the experiment. For extrapolation to ``Phase IV'' of the experiment, it is slightly more challenging as the superconductor material in the detectors will change, in addition to the inclusion of more data using a larger number of pixels and days of running. To model the improved materials, we perform a simplified sterile neutrino analysis to compare with Ref.~\cite{Acero:2022wqg} and find the scale factor that would obtain the preliminary reach for phase 4 (around $2\times 10^3$). With this factor in hand, we also scale the number of events to 128 pixels over 100 days. 

Finally, it is worth noting that the decay processes to the excited state involve the subsequent emission of a photon that can be used to tag the excited processes~\cite{Bray:2024brg}. This can be useful in reducing the tails around the ground state peaks, which could improve the reach for not only neutrinophilic scalars but also the search for sterile neutrinos. For projections that include coincidence tagging of the $477~\mathrm{keV}$ $\gamma$ ray, an assumed excited-state tagging efficiency of order $90\%$ is used to illustrate the potential improvement in background discrimination, noting that the achievable efficiency depends on detector geometry and $\gamma$-ray detection performance.

We show our estimated limits on $\lambda_{e\ell}$ as a function of $m_\phi$ in Fig.~\ref{fig:beestbound}. We note that the excluded region is in tension with a standard cosmological history (just as sterile neutrinos with mass below $\sim{\rm MeV}$ and $U^2\gtrsim10^{-10}$ are). However, these bounds can be weakened in nonstandard cosmological scenarios, such as ``mass varying neutrinos''~\cite{Fardon:2003eh,Fardon:2005wc}, where the finite cosmological density of light neutrinos can keep a light scalar field that couples directly to sterile neutrinos, $A$, away from the true minimum of its potential. An interaction of the form $A^2|\phi|^2$ could allow for $m_\phi$ to be large at early times when the neutrino number density was higher, suppressing the production of $\phi$ in the early Universe. We emphasize that it is particularly necessary to test the possibility of such nonstandard cosmologies with laboratory experiments, especially given the tensions in cosmological observations that could be alleviated by stronger-than-weak neutrino self-interactions. Model building this scenario is the subject of ongoing work.

\section{Conclusion}
\label{sec:conc}

In this work, we investigate how light neutrinophilic scalars affect low-energy charged-current processes, specifically pion decays and nuclear electron capture. In the presence of such a scalar, both processes acquire a new three-body channel that generates a characteristic low-energy tail in the visible energy spectrum. We show that these processes suffer divergences when $m_\phi\to0$ at tree-level, which are tamed once one-loop corrections are included.

For pion decay, we focused on the electron channel. This channel provides the strongest sensitivity for PIONEER, since the two-body SM mode is helicity-suppressed, allowing the three-body contribution to stand out more prominently. We showed that PIONEER’s energy resolution and statistics make it capable of resolving the low energy tail over a wide range of mediator masses. This translates into sensitivity to couplings well beyond existing laboratory constraints. Note that the same analysis can be carried out for the muon channel. Of course, a positive BSM signal in the $\pi^+\to e^+ + {\rm inv.}$ channel would not by itself be proof of a neutrinophilic boson as it could also be associated with a boson that couples to charged leptons or even quarks; see, e.g., Ref.~\cite{wgpioneer} for a study the sensitivity of PIONEER to wide classes of invisibly-decaying particles with couplings to neutrinos and/or other SM fermions. Determining precisely which model gives rise to an excess above SM expectation would involve detailed studies of the $e^+$ spectrum as well as related signals in other experiments.

We also analyzed nuclear electron-capture decays, using BeEST as a concrete experimental example. While the current BeEST configuration is not competitive with pion measurements (particularly because the reach for scalar emission does not weaken at low masses, unlike sterile neutrino searches), the experimental setup is intrinsically scalable, with potential to explore different nuclei with different $Q$ values. Other EC experiments targeting neutrino properties, such as HUNTER~\cite{Smith:2016vku,Martoff:2021vxp} or HOLMES~\cite{HOLMES:2016spk,HOLMES:2025qzi}, could potentially look for neutrinophilic scalars. Moreover, EC represents only one member of a larger class of capture processes. Muon capture experiments using muonic atoms can access heavier states~\cite{McKeen:2010rx,Fox:2024kda}, comparable in mass to those available in pion decay. Fully exploiting this pathway could require high precision calculations of nuclear matrix elements, as in, e.g.,~\cite{Jokiniemi:2024zdl}.

\acknowledgements
We thank Douglas Bryman, Benjamin Davis-Purcell, Stephan Friedrich, Michael Gennari, Emma Klemets, Kyle Leach, Chloé Malbrunot, Richard Mischke, Toshio Numao, Robert Shrock, Patrick Schwendimann,  and  Bob Velghe for helpful discussions and comments on our manuscript, as well as the authors of~\cite{wgpioneer} for making their work available to us before publication and for coordinating submission to the arXiv. We also thank Danaan Cordoni-Jordan for collaboration in the early stages of this work. This work is supported by Discovery Grants from the Natural Sciences and Engineering Research Council of Canada (NSERC). TRIUMF receives federal funding via a contribution agreement with the National Research Council (NRC) of Canada.

\bibliography{bibref}

\end{document}